\documentclass[prb, preprint,amsmath, amssymb,]{revtex4-1}

\usepackage{graphicx}
\usepackage{dcolumn}
\usepackage{bm}

\begin{document}

\preprint{jcp}

\title[Pattern formation in bistable media]
{Pattern formation controlled by time-delayed feedback in bistable media}

\author{Ya-feng He$^{1,2}$}
\email[Email:]{heyf@hbu.edu.cn}
\author{Bao-quan Ai$^{1,3}$}
\author{Bambi Hu$^{1,4}$}

\affiliation{$^1$Centre for Nonlinear Studies, The Beijing-Hong Kong-Singapore Joint Centre for Nonlinear
and Complex Systems (Hong Kong), Hong Kong Baptist University, Kowloon Tong, Hong Kong, China\\
$^2$College of Physics Science and Technology, Hebei University, Baoding 071002, China\\
$^3$Laboratory of Quantum Information Technology, ICMP and SPTE, South China Normal University, Guangzhou 510006, China
\\$^4$Department of Physics, University of Houston, Houston, TX 77204-5005, USA.}

\date{\today}
\begin{abstract}
Effects of time-delayed feedback on pattern formation are studied both numerically and theoretically in a bistable reaction-diffusion model. The time-delayed feedback applied to the activator and/or the inhibitor alters the behavior of the Nonequilibrium Ising-Bloch (NIB) bifurcation. If the intensities of the feedbacks applied to the two species are identical, only the velocities of Bloch fronts are changed. If the intensities are different, both the critical point of the NIB bifurcation and the threshold of stability of front to transverse perturbations are changed. The effect of time-delayed feedback on the activator opposes the effect of time-delayed feedback on the inhibitor. When the time-delayed feedback is applied individually to one of the species, positive and negative feedbacks make the bifurcation point shift to different directions. The time-delayed feedback provides a flexible way to control the NIB bifurcation and the pattern formation.
\end{abstract}

\maketitle

\section{INTRODUCTION}

\indent Pattern formation has been of great interest in a variety of chemical, physical, and biological contexts. \cite{Cross, Koch, Judit, He, Epstein} A chemical front (interface) which connects two different states of system, such as the excited and recovery states in excitable media, or the two stable states in bistable media, plays an essential role on the pattern formation. Many literatures focus on the dynamics of the front. \cite{Kothe, Zykov, Hagberg1, Marts, Bar1, Haas, Lee1, Hagberg2, Bar2, Goldstein} Of particular interest is the front controlled by Nonequilibrium Ising-Bloch (NIB) bifurcation in the bistable media. The NIB bifurcation describes a pitchfork bifurcation at which a stationary Ising front becomes unstable and a couple of counterpropagating Bloch fronts appear. In bistable Ferrocyanide-Iodate-Sulfite reactions, spirals, oscillating spots, and labyrinthine patterns have been observed. \cite{LiGe, Szalai, Lee1, Lee2} The spirals occur in the Bloch region beyond the Nonequilibrium Ising-Bloch bifurcation. As a sparse spiral, it results from an axisymmetry breaking of a shrinking ring. The oscillating spots appear near but before the NIB bifurcation. The labyrinthine pattern originates from transverse instability of a chemical front in Ising region. Similar patterns were also observed by Szalai and De Kepper. \cite{Szalai} These patterns observed in bistable Ferrocyanide-Iodate-Sulfite reactions can be explained successfully in terms of a NIB bifurcation in a generic FitzHugh-Nagumo model. \cite{Hagberg1, Hagberg0, Hagberg3}

\indent Controlling the pattern formation is an important issue for the study of self-organization phenomena far away from thermodynamic equilibrium. Recently, time-delayed feedback, firstly presented by Ott $et$ $al$ to control the chaotic behavior of a deterministic system, \cite{Ott} has been used to control the pattern formation successfully. It can control the tip trajectories of spirals in a light-sensitive Belousov-Zhabotinsky reaction. \cite{Kheowan} A global feedback can either stabilize the rigid rotation of a spiral or completely destroy spiral and suppress self-sustained activity in a confined domain of excitable medium. \cite{Zykov2} The spontaneous suppression of spiral turbulence based on feedback has been studied experimentally in a light-sensitive Belousov-Zhabotinsky reaction and numerically in a modified FitzHugh-Nagumo model. \cite{Guo} With the global feedback one can manipulate the competition between patterns with different symmetries (hexagons and rolls). \cite{Stanton} In a delayed optical system, resonant Hopf triads lead to drifting rhombic and hexagonal patterns. \cite{Logvin} Near the codimension-two bifurcation points, the time delay can result in a transition between Turing and Hopf instabilities. \cite{Sen, Li}

\indent Most of the studies of the effects of time-delayed feedback on the pattern formation focus on the dynamics of patterns in excitable and oscillatory media. \cite{Zykov2, Gassel, Guo, Balanov, Li2, Kim, Kheowan, Schneider, Sen, Li} How about the effects of time-delayed feedback on controlling pattern formation in bistable media? In this work, we study the role of time-delayed feedback played on controlling the NIB bifurcation in a bistable FitzHugh-Nagumo model. We focus on the investigation of the NIB bifurcation and the transverse instability of front when applying time delay. The underlying mechanism of successful control is analyzed.

\section{BISTABLE MODEL}

\indent The bistable media are described by a FitzHugh-Nagumo model:
\begin{eqnarray}
  u_t &=& u -u^3 - v + \nabla^2 u +F,\\
  v_t &=& \varepsilon(u-a_{1}v-a_{0})+\delta \nabla^2 v +G,
\end{eqnarray}
where the time delay is applied with the forms:
\begin{eqnarray}
  F=g_{u}(u(t-\tau)-u(t)),\\
  G=g_{v}(v(t-\tau)-v(t)),
\end{eqnarray}
here, the variables $u$ and $v$ represent the concentrations of the activator and inhibitor, respectively, and $\delta$ denotes the ratio of their diffusion coefficients. $\tau$ indicates the delayed time. $g_{u}$ and $g_{v}$ are the feedback intensities of variable $u$ and $v$, respectively. The small value $\varepsilon$ characterizes the time scales of the two variables, where $v$ remains approximately constant $v_{f}$ on the length scale over which $u$ varies.  The system described by Eqs.(1) and (2) can be either of excitable, Turing-Hopf, or bistable type. In this paper the parameter $a_{1}$ is chosen such that the system is bistable. The two stationary and uniform stable states are indicated by an up state ($u_{+}$,$v_{+}$) and a down state ($u_{-}$,$v_{-}$), respectively. The parameter $a_{0}$ represents the symmetry of the system. In the following, we only consider the case that the system is symmetric, i.e. $a_{0}$$=$$0$, ($u_{+}$,$v_{+}$)=-($u_{-}$,$v_{-}$). A front connects the two stable states smoothly. It can be either traveling (Bloch front) or stationary (Ising) which is determined by the control parameters.

\indent Because pattern formations in bistable media are sensitive to the initial and boundary conditions, we adopt fixed initial conditions during the numerical simulations. In the one-dimensional case ($200$ grids, using Euler method), we focus on the traveling wave with an initial condition as shown in Fig.1 (a). In the two-dimensional case ($200$$\times$$200$ grids, using Peaceman-Rachford alternating-direction implicit scheme), we mainly concentrate on the spiral wave and labyrinthine pattern with an asymmetrical initial condition indicated in Fig. 5 (a). The boundary conditions are taken to be periodic and no-flux in one and two dimensions, respectively. The space step is $dx$$=$$dy$$=$$1.0$ length unit and the time step is $dt$$=$$0.05$ time unit in both cases.

\section{RESULTS AND DISCUSSION}
\subsection{Front bifurcation in one dimension}

\indent Firstly, we study numerically the front bifurcation in one dimension. Without the time delay, the system follows a NIB bifurcation upon decreasing $\varepsilon$, which leads to a transition from stationary Ising front to a couple of counterpropagating Bloch fronts. Figure 1 (b) shows the time evolution of the two Bloch fronts with the parameters deep into Bloch region. The traveling wave propagates at constant speed. Now, we apply a time delay with the forms in Eqs. (3) and (4), but still keep the control parameters the same as those in Fig. 1 (b). When $\tau$$=$$0.2$, $g_{u}$$=$$0.9$, $g_{v}$$=$$0.1$, the original Bloch fronts slow down and finally stop at some place as shown in Fig. 1 (c). This means that the time delay alters the point of NIB bifurcation, which results in a transition from initial Bloch fronts into Ising fronts.

\begin{figure}[htbp]
  \begin{center}\includegraphics[width=8cm,height=6cm]{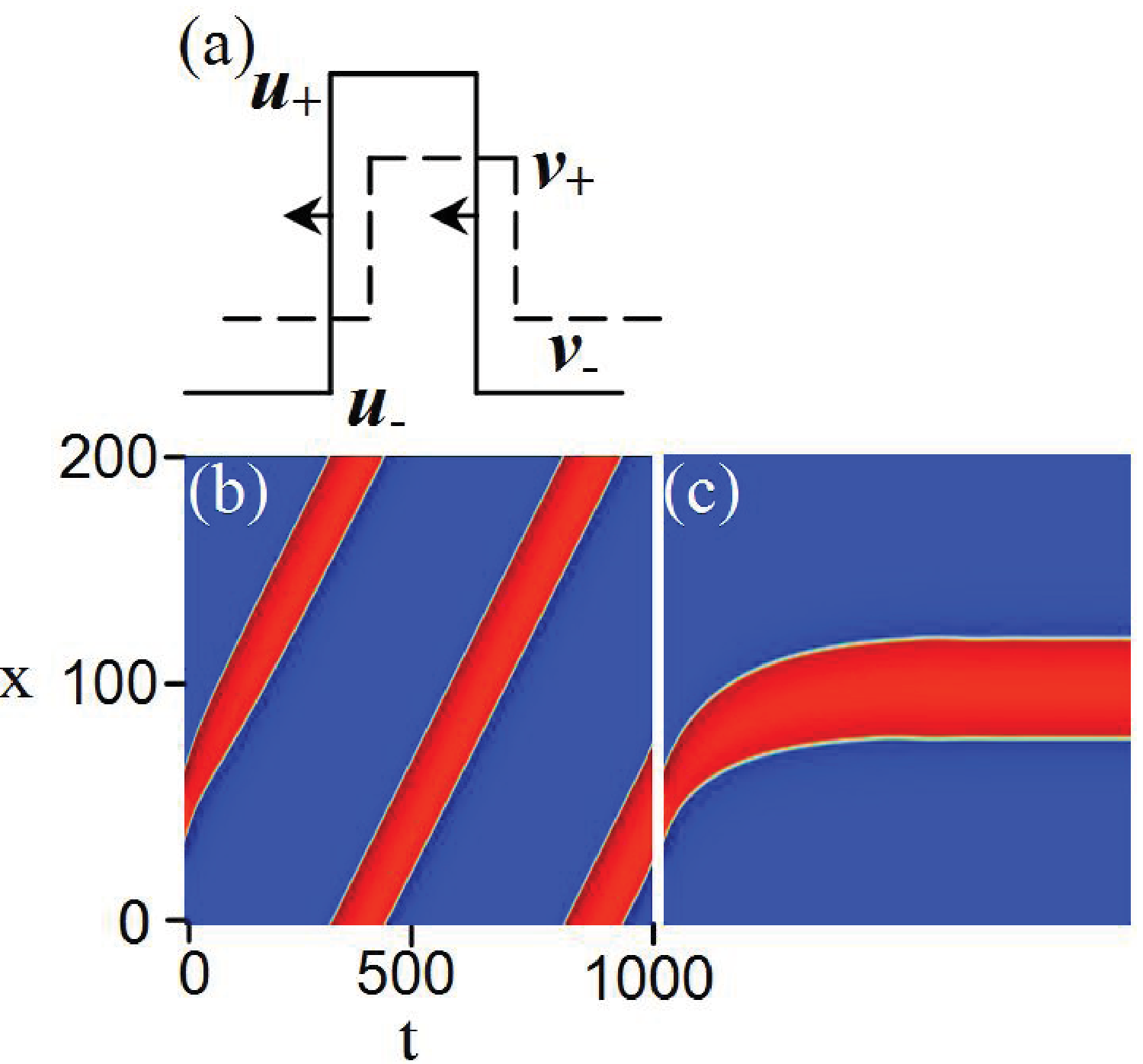}
  \caption{Time evolution of a front and a back. (a) Initial condition; (b) $\tau$$=$$0.0$; (c) $\tau$$=$$0.2$, $g_{u}$$=$$0.9$, $g_{v}$$=$$0.1$. The other control parameters are: $a_{1}$$=$$2.0$, $\delta$$=$$2.0$, and $\varepsilon$$=$$0.03$. The time-space span in (c) is the same as that in (b).}\label{1}
  \end{center}
\end{figure}

\indent We obtain numerically the dependence of the front velocity on the parameter $\varepsilon$ as shown in Fig. 2, in order to investigate the front bifurcation. In the absence of time-delayed feedback, the NIB bifurcation occurs at $\varepsilon$$=$$0.036$ and it is a pitchfork bifurcation as indicated by the solid circle in Fig. 2 (a) and (b). Then, we apply positive feedback to the system with the delayed time $\tau$$=$$0.2$ and the identical feedback intensities $g_{u}$$=$$g_{v}$$=$$0.9$. From the numerical results represented by the empty circle in Fig. 2 (a), it is shown that the opening angle of the pitchfork shrinks. The larger the delayed time $\tau$ is, the smaller the opening angle becomes. So, by using time delay one can reduce the front velocity. It is interesting that under the feedback with identical intensities the critical point of the NIB bifurcation doesn't vary. If we apply the time delay with different feedback intensities the situations become much different. On one hand, if the feedback intensity $g_{u}$ is larger than $g_{v}$, the bifurcation point shifts to the left and the opening angle of the pitchfork reduces to some extent as shown by the solid square in Fig. 2 (a). The front speed in this case is smaller than that in the absence of time delay. This case can induce a transformation from Bloch fronts to Ising fronts as shown in Fig. 1. On the other hand, if the feedback intensity $g_{u}$ is smaller than $g_{v}$, the bifurcation point shifts to the right, and the opening angle reduce as indicated by the empty square in Fig. 2 (a). The empty square line may intersect with the solid circle line (without time delay) for small $\tau$. On the right of the crossover point the front speed with time delay is larger than that without time delay. It is clearly that under the time delay with $g_{u}$$<$$g_{v}$, the initial Ising front loses stability to evolve into Bloch front. In general, upon increasing the ratio of the feedback intensities $g_{u}/g_{v}$, the bifurcation point shifts to the left gradually.

\begin{figure}[htbp]
  \begin{center}\includegraphics[width=8cm,height=10cm]{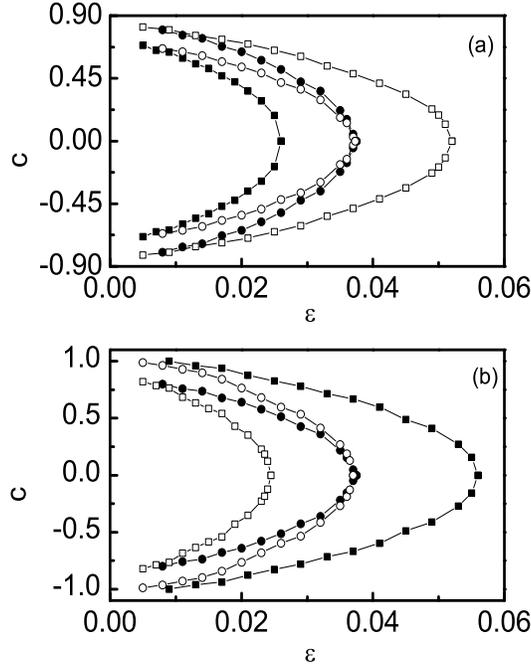}
  \caption{Dependence of the front velocity on the parameter $\varepsilon$. (a) Positive feedback, solid circle: $\tau$$=$$0.0$; empty circle: $\tau$$=$$0.2$, $g_{u}$$=$$0.9$, $g_{v}$$=$$0.9$; solid square: $\tau$$=$$0.2$, $g_{u}$$=$$0.9$, $g_{v}$$=$$0.1$; empty square : $\tau$$=$$0.2$, $g_{u}$$=$$0.1$, $g_{v}$$=$$0.9$. (b) Negative feedback, solid circle: $\tau$$=$$0.0$; empty circle: $\tau$$=$$0.2$, $g_{u}$$=$$-0.9$, $g_{v}$$=$$-0.9$; solid square: $\tau$$=$$0.2$, $g_{u}$$=$$-0.9$, $g_{v}$$=$$-0.1$; empty square : $\tau$$=$$0.2$, $g_{u}$$=$$-0.1$, $g_{v}$$=$$-0.9$. The other parameters are: $\delta$$=$$2.0$, and $a_{1}$$=$$2.0$.}\label{2}
\end{center}
\end{figure}

\indent In the negative feedback case, $g_{u}$$<$$0$, $g_{v}$$<$$0$, the opposite is true. The opening angle enlarges, leading to the increasing of the front speed. Upon increasing the ratio of the feedback intensities $g_{u}/g_{v}$, the bifurcation point shifts to the right gradually as shown in Fig. 2 (b).

\indent It shows that the effect of time-delayed feedback on the first variable opposes that on the second variable. There exists competition between the two feedbacks on controlling the NIB bifurcation. If the feedback intensities acting on the two variables are identical, the NIB bifurcation point does not affected by the time-delayed feedback as shown above.

\indent If the feedback is applied individually, such that $g_{u}$$=$$0$, or $g_{v}$$=$$0$, we can still realize the shift of the critical point of NIB bifurcation. Increasing the feedback $g_{u}$ ($g_{v}$$=$$0$), for instance from negative to positive values, the bifurcation point shifts from right to left gradually. On the contrary, if increasing the feedback $g_{v}$ ($g_{u}$$=$$0$) from negative to positive values, the bifurcation point shifts from left to right. It shows that the effect of the time delay with positive feedback on the variables opposes the effect of time delay with negative feedback on the variables. Therefore, by using time delay with appropriate forms one can control the front bifurcation efficiently.

\indent In the absence of the time-delayed feedback the front bifurcation in one dimension is determined by the relation between the front velocity and the parameter $\varepsilon$, \cite{Hagberg2,Hagberg3}

\begin{equation}\label{}
c=\frac{3c}{\sqrt{2}q^{2}\big(c^{2}+4\varepsilon\delta q^{2})^{\frac{1}{2}}},
\end{equation}
where $q=\sqrt{a_{1}+\frac{1}{2}}$. In Eqs. (3) and (4), if the delayed time $\tau$ is small, we can expand $u(t-\tau)$ and $v(t-\tau)$ as,

\begin{eqnarray}
  u(t-\tau) &=& u(t) -\tau \frac{\partial u(t)}{\partial t},  \\
  v(t-\tau) &=& v(t) -\tau \frac{\partial v(t)}{\partial t}.
\end{eqnarray}
So, we can obtain:

\begin{eqnarray}
  (1+\tau g_{u})u_t &=& u -u^3 - v + \nabla^2 u,  \\
  (1+\tau g_{v})v_t &=& \varepsilon(u-a_{1}v)+\delta \nabla^2 v.
\end{eqnarray}
It shows that the time delay affects the temporal scales of the variables. We use the singular perturbation analysis to study the front bifurcation assuming $\varepsilon/\delta$$\ll$$1$. Following Eq. (5) it is straightforward to obtain the implicit expression of the front velocity:

\begin{equation}\label{}
c(1+\tau g_{u})=\frac{3c(1+\tau g_{v})}{\sqrt{2}q^{2}[c^{2}(1+\tau g_{v})^{2}+4\varepsilon\delta q^{2}]^{\frac{1}{2}}}.
\end{equation}
Obviously, the feedback parameters $\tau$, $g_{u}$ and $g_{v}$ affect the behaviors of the front bifurcation.

\begin{figure}[htbp]
  \begin{center}\includegraphics[width=8cm,height=10cm]{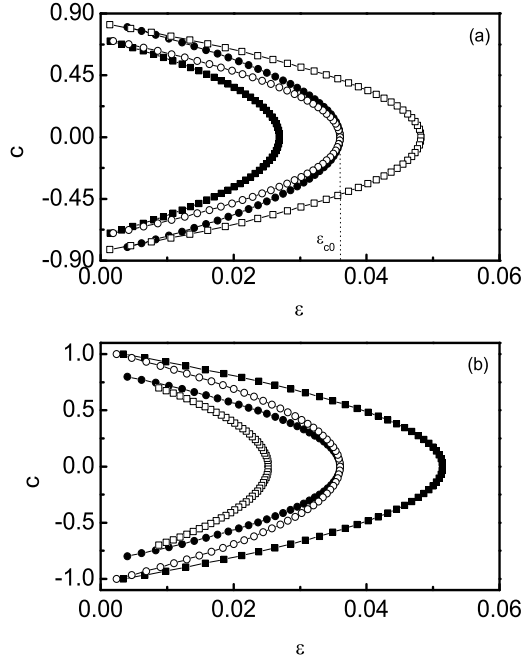}
  \caption{Plots of Eq. (10) in the ($c$, $\varepsilon$) plane. The control parameters and the notations correspond to those in Fig. 2}
  \end{center}
\end{figure}

\indent Figure 3 plots the dependence of the front velocity on the parameter $\varepsilon$ based on the Eq. (10). In the case of identical feedbacks, such that $g_{u}$$=$$g_{v}$, the velocity of Bloch fronts can be rescaled. If $g_{u}$$=$$g_{v}$$>$$0$, the front velocity decreases that leading to the reduction of the opening angle of the pitchfork as shown in Fig. 3 (a). On the contrary, if $g_{u}$$=$$g_{v}$$<$$0$, the final front velocity increases, which leads to the increasing of the opening angle as indicated in Fig. 3 (b).

\indent Next, we focus on the critical point of front bifurcation, at which $c$$=$$0$. Thus, Eq. (10) can be reduced to:

\begin{equation}\label{}
\varepsilon_{c}=\frac{9}{8\delta q^{6}}\left(\frac{1+\tau g_{v}}{1+\tau g_{u}}\right)^{2}.
\end{equation}
In the absence of the time delay, such that $\tau$$=$$0$, we denote the critical value of the front bifurcation by $\varepsilon_{c0}$. It can be found from Eq. (11) that if $g_{u}$$>$$g_{v}$, the critical value shifts to the left. On the contrary, if $g_{u}$$<$$g_{v}$, it shifts to the right. There exists the competition between the feedback $g_{u}$ and $g_{v}$ on controlling the front bifurcation. The corresponding bifurcation diagrams are shown in Fig. 3 (a) and (b), respectively. It can explain well the numerical results in Fig. 2. The front bifurcations, both without and with the time delay, are plotted in the $\delta-\varepsilon$ plane as indicated by the thick dash line $\delta_{F}$ and the thick dash dot line $\delta_{FD}$ in Fig. 7.

\begin{figure}[htbp]
  \begin{center}\includegraphics[width=8cm,height=8cm]{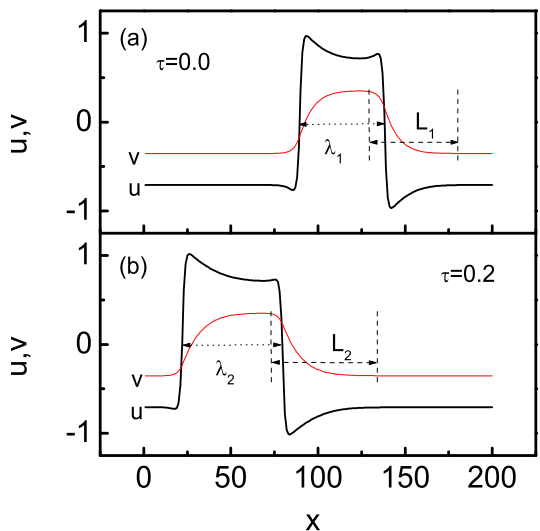}
  \caption{Profiles of Bloch fronts without (a) and with (b) the time-delayed feedback. The tailing of Bloch front $L_{1}$ in (a) is smaller than $L_{2}$ in (b). The width of up state $\lambda_{2}$$>$$\lambda_{1}$. The feedback parameters in (b) are: $\tau$$=$$0.2$, $g_{u}$$=$$0.1$, $g_{v}$$=$$0.9$. The other parameters are: $\delta$$=$$2.0$, and $a_{1}$$=$$2.0$.}
  \end{center}
\end{figure}

\indent Applying the time-delayed feedback to the variables is equivalent in some sense to changing their diffusion coefficients. For example, applying the time-delayed feedback with $|1+\tau g_{u}|$$<$$|1+\tau g_{v}|$ is equivalent to slowing the diffusion of the inhibitor [see Eq. (11)], therefore, increasing the velocity of Bloch front. Because the width of Bloch front is in inverse proportion to the diffusion coefficient $\delta$, applying the time delay with $|1+\tau g_{u}|$$<$$|1+\tau g_{v}|$ widens the Bloch front. Figure 4 shows the profiles of Bloch fronts without and with the time-delayed feedback. Obviously, the tailing $L_{2}$ in (b) is wider than $L_{1}$ in (a) and the Bloch front is widen by the time-delayed feedback. Therefore, the width of the up state is widened, $\lambda_{2}$$>$$\lambda_{1}$, resulting in an increase in the wavelength of the spiral wave in two dimension. On the contrary, if $|1+\tau g_{u}|$$>$$|1+\tau g_{v}|$, the opposite is true.

\indent We want to point out here that with extensive numerical simulation the above results are still correct when long delays are applied. The feedbacks $g_{u}$ and  $g_{v}$ can be applied either individually or simultaneously, which depends on their values. For example, if individual feedback $g_{v}$ with large delay is applied, $\tau$$=$$20$, $g_{u}$$=$$0.0$, and $g_{v}$$=$$0.1$, the bifurcation point still shifts right, which is equivalent to the manipulation $\tau$$=$$0.2$, $g_{u}$$=$$0.0$, and $g_{v}$$=$$10$. The larger the product $\tau$$\ast$$g_{v}$ is, the farther the bifurcation point deviates from the critical point $\varepsilon_{c0}$. This is because that the delayed time $\tau$ and the feedback intensity $g_{u}$ ($g_{v}$) are coupled together as indicated in the derived Eqs. (8) and (9). Obviously, the action of long delay with weak feedback intensity is equivalent to that of short delay with strong feedback intensity. This provides guidance in practical application. In order to keep the applicability of Taylor expansion on deriving the Eqs. (8) and (9) and the consistency between the analytical and numerical results, we used small delays throughout the manuscript.

\subsection{Transverse instability of front in two dimensions}

\indent In one dimensional case we concentrate on the front bifurcation by analyzing the relation between the velocity of front and the parameter $\varepsilon$. A planar front could become curve in two dimensions. It is necessary to consider further the stability of a planar front to transverse perturbation, i.e. the transverse instability of planar front. In this section, we firstly obtain patterns deep into the Bloch and Ising regions, and near the NIB bifurcation point without time delay. Then we study the effects of time delay on the transverse instability of front. Our emphasis is on controlling the transverse instability of front by applying appropriate time delay.

\begin{figure}[htbp]
  \begin{center}\includegraphics[width=8cm,height=5cm]{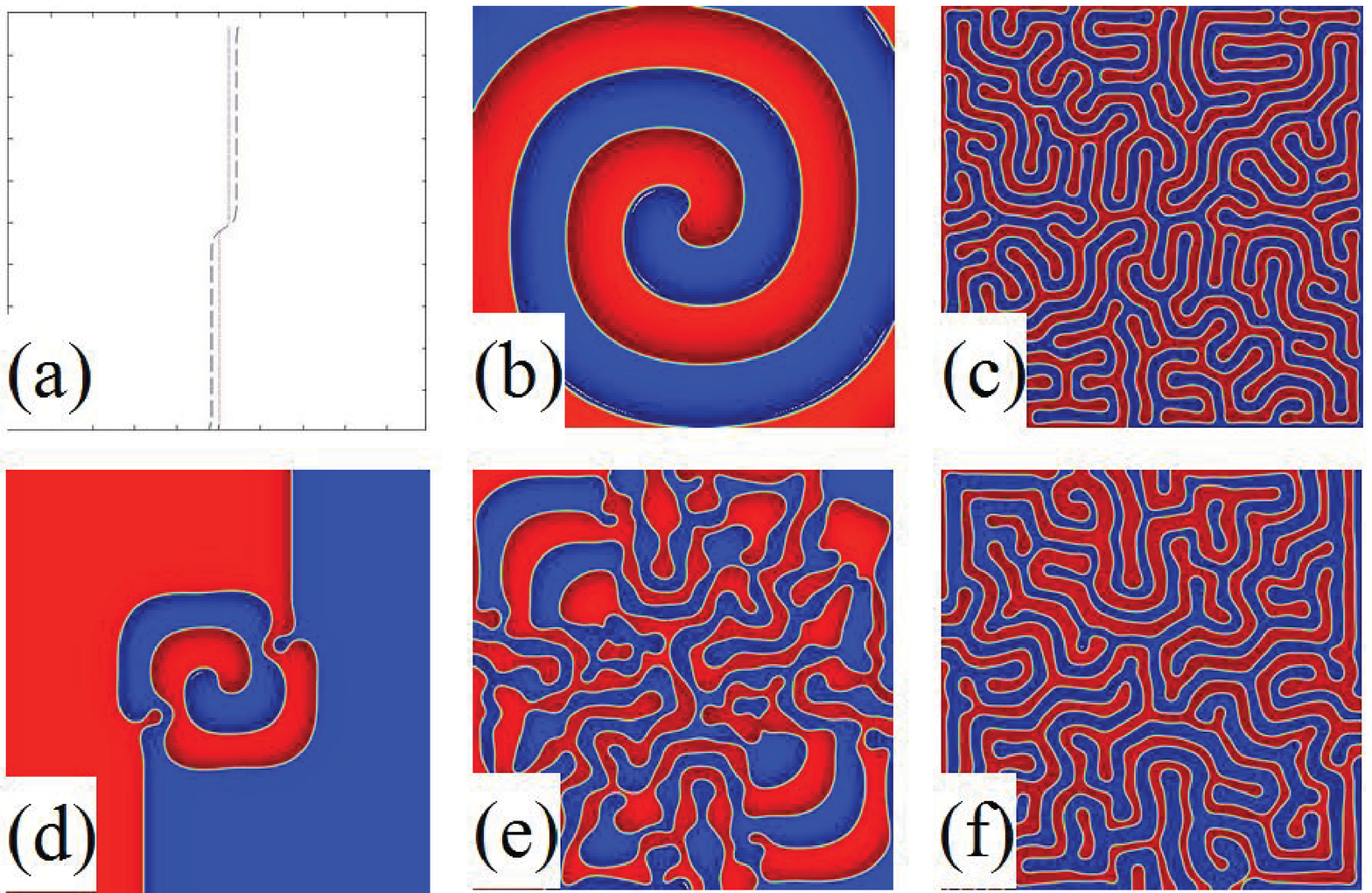}
  \caption{Numerical simulations of the symmetrically bistable model. (a) Initial condition; The dash line (dot line) represents an interface of variable $u$ ($v$), which separates the up state (the left part) and the down state (the right part). (b) Spiral wave, $\varepsilon$$=$$0.02$; (c) Stationary Labyrinthine, $\varepsilon$$=$$0.05$;  (d)-(f) are the snapshots of the evolvement of breathing labyrinthine at $\varepsilon$$=$$0.036$, $t$$=$$250$, $650$, $2000$ time units. The other parameters are: $\delta$$=$$2.0$, and $a_{1}$$=$$2.0$. Grid size: $200$$\times$$200$ space units.}\label{4}
  \end{center}
\end{figure}

\indent In order to illustrate and compare the results clearly, we use the same initial condition as shown in figure 5 (a). The intersection point of the contours of $u$ and $v$ servers as an initial tip for the spiral formation. From Eq. (10) it can be seen that the parameters $\varepsilon$ and $\delta$ are coupled together. In the following, for simplicity, we keep the parameter $\delta$ constant.

\indent Deep into the Bloch region, a couple of Bloch fronts counterpropagate and form spiral wave as indicated in Fig. 5 (b). The fronts are stable to transverse perturbations. In the set of present parameters the obtained spiral is a dense spiral (the up state and the down state are symmetric except an angle separation of $\pi$). Deep into the Ising region, starting from the initial condition [Fig. 5 (a)], the part near the domain center firstly evolves into a spiral head. Then, the part behind the spiral head undergoes transverse instability and the fronts interplay with each other, which resulting in a stationary labyrinthine pattern finally [Fig. 5 (c)]. In the stationary labyrinthine pattern the up states keep interconnection and own identical widths. This process is similar with the observation in Refs. 15, 20. Near the NIB bifurcation, the situation becomes more complex, where we observe a breathing labyrinthine pattern. Fig. 5 (d)-(f) show three snapshots of the evolvement of breathing labyrinthine at $t$$=$$250$, $650$, $2000$ time units. In this case, the up states can breakdown and reconnect. Together with repulsive interaction between fronts, the widths of the up states increase and decrease periodically, leading to the formation of a breathing labyrinthine pattern. The most difference between the breathing labyrinthine and the stationary labyrinthine is that in breathing labyrinthine case the up state does not interconnect entirely and its width changes periodically. The present dynamics is similar with that of oscillatory spots. \cite{Lee2}

\begin{figure}[htbp]
  \begin{center}\includegraphics[width=8cm,height=6cm]{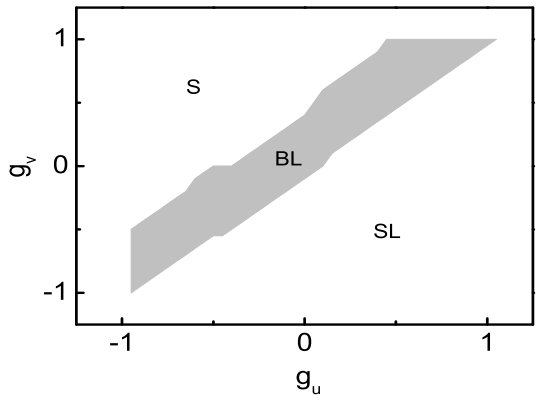}
  \caption{Phase diagram spanned by the feedback intensities $g_{u}$ and $g_{v}$. S--Spiral; BL--Breathing Labyrinthine; SL--Stationary Labyrinthine. The other parameters are: $\delta$$=$$2.0$, $a_{1}$$=$$2.0$, $\varepsilon$$=$$0.036$, and $\tau$$=$$0.2$.}\label{5}
  \end{center}
\end{figure}

\indent In order to illustrate clearly the transformation between various patterns controlled by the time delay, we still use the above individual parameter sets and the initial condition [Fig. 5 (a)]. Starting from the parameters set in Fig. 5 (f), in which a breathing labyrinthine forms in the absence of the time delay, if the feedback intensity $g_{u}$$>$$g_{v}$ the given initial condition evolves into stationary labyrinthine pattern. However, if $g_{u}$$<$$g_{v}$ it transits into the spiral pattern. So, the time delay can alter the critical value of transverse instability of planar front (see the analysis below, Fig. 7). When applying the time delay to the system with the parameters in Fig. 5 (b) and the same initial condition [Fig. 5 (a)], upon increasing the ratio of $g_{u}/g_{v}$, it will develop into breathing labyrinthine and stationary labyrinthine patterns successively. Similarly, if decreasing the ratio of $g_{u}/g_{v}$ with the parameters as in Fig. 5 (c), breathing labyrinthine and spiral patterns form in sequence. Figure 6 shows a phase diagram spanned by the feedback intensities $g_{u}$ and $g_{v}$, in which the gray region represents the breathing labyrinthine pattern. It should be mentioned that the boundary between spiral patterns and breathing labyrinthine patterns is not sharp because near this boundary the arm of the spiral far away the tip could reflect upon touching the domain boundary which leading to the breakdown of the arm. Here, we plot the boundary at which perfect spirals could form. The wavelength of spiral can be adjusted by varying the feedback parameters, as we have depicted above in the one dimensional case. The spiral period is around 160 time units. So, the time delay is still applicable for controlling spiral patterns. Therefore, by varying the ratio $g_{u}/g_{v}$ one can realize the control of transverse instability of planar front.

\begin{figure}[htbp]
  \begin{center}\includegraphics[width=8cm,height=6cm]{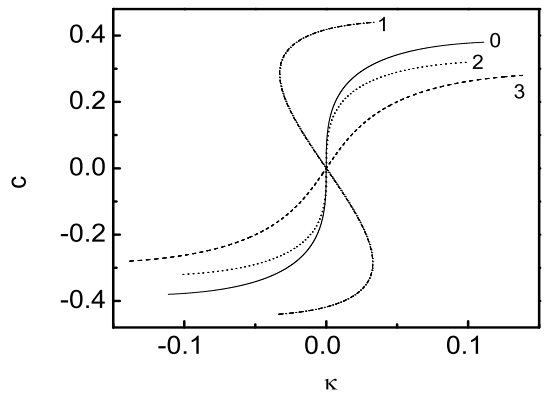}
  \caption{Dependence of the front velocity on the curvature. Solid line (0): without time delay. Dash dot line (1): $\tau$$=$$0.2$, $g_{u}$$=$$0.1$, $g_{v}$$=$$0.9$; Dotted line (2): $\tau$$=$$0.2$, $g_{u}$$=$$0.9$, $g_{v}$$=$$0.9$; Dash line (3): $\tau$$=$$0.2$, $g_{u}$$=$$0.9$, $g_{v}$$=$$0.1$. The other parameters are: $\delta$$=$$2.0$, $a_{1}$$=$$2.0$, $\varepsilon$$=$$0.036$.}\label{6}
  \end{center}
\end{figure}

\indent In two dimensions the front velocity is modified by the curvature of front. We should consider the transverse instability of planar front beside the NIB bifurcation. Here, we use the algorithm in Ref. 20 to analyze the transverse instability of both Ising and Bloch fronts. Under the modification by curvature, Eq. (10) can be written as:

\begin{equation}\label{}
c_{r}(1+\tau g_{u})+\kappa=\frac{3(c_{r}(1+\tau g_{v})+\delta\kappa)}{\sqrt{2}q^{2}[(c_{r}(1+\tau g_{v})+\delta\kappa)^{2}+4\varepsilon\delta q^{2}]^{\frac{1}{2}}},
\end{equation}
here, $c_{r}$ is the normal velocity, and $\kappa$ presents the curvature. Figure 7 shows a velocity-curvature relation without and with the time delay. The solid line (0) represents the breathing labyrinthine pattern at NIB bifurcation without time delay [Fig. 5 (f)]. At the center of the plot, the slope of the curve indicates critical stability. If applying time delay with identical intensities, such that $g_{u}$$=$$g_{v}$, the velocity changes, but the slope of the curve at the center still keeps constant [dotted line (2)]. The stability of front to perturbation hardly varies. So, one can still observe breathing labyrinthine pattern. When the feedback intensities $g_{u}$$>$$g_{v}$, the above slope is positive [dash line (3)]. A front becomes unstable to perturbation, and it finally evolves into stationary labyrinthine pattern. On the contrary, if $g_{u}$$<$$g_{v}$, the mentioned slope becomes negative [dash dot line (1)], and a front keeps stable upon suffering perturbation. We can obtain spiral pattern as shown above.

\begin{figure}[htbp]
  \begin{center}\includegraphics[width=8cm,height=10cm]{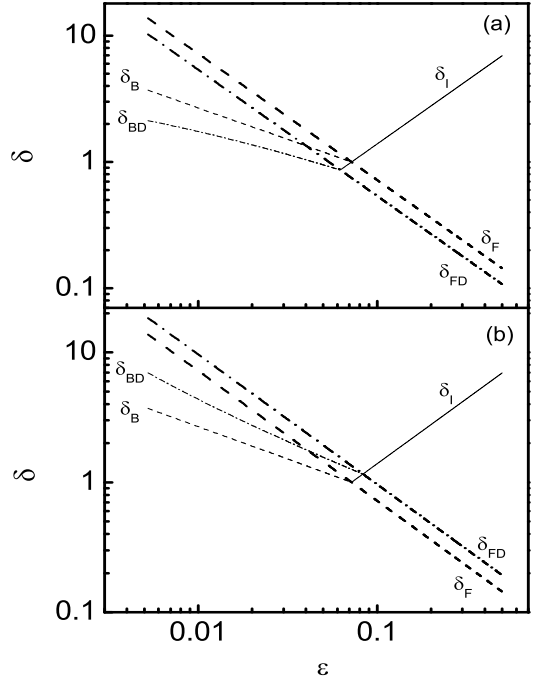}
  \caption{Phase diagram for the NIB bifurcation and the transverse instability. The solid line $\delta_{I}$ indicates the Ising front. The thick dash line $\delta_{F}$ and dash dot line $\delta_{FD}$ present the front bifurcations without and with time delay, respectively. The thin dash line $\delta_{B}$ and dash dot line $\delta_{BD}$ present the Bloch front without and with time delay, respectively. The feedback intensity in (a) $g_{u}$$=$$0.9$, $g_{v}$$=$$0.1$; (b) $g_{u}$$=$$0.1$, $g_{v}$$=$$0.9$. The other parameters are: $a_{1}$$=$$2.0$, and $\tau$$=$$0.2$.}\label{7}
\end{center}
\end{figure}

\indent We now analyze further the stabilities of both Bloch and Ising fronts to perturbation when applying the time delay. If the curvature is small, the normal velocity $c_{r}$ can be replaced by $c_{r}$$=$$c_{0}$$-$$d\kappa$, in which $c_{0}$ indicates the velocity of planar front. Here, the reduced parameter $d$ is not anymore a simple diffusion coefficient of activator as in excitable system. \cite{Hagberg3} Its sign determines the stability of a front to transverse perturbations. Inserting $c_{r}$ into Eq. (12) and taking Taylor expansion, we can obtain the implicit expression about $d$:

\begin{equation}\label{}
 1-d(1+\tau g_{u})=\frac{3\big(\delta-d(1+\tau g_{v}))}{\sqrt{2} q^{2} [c_{0}^{2}(1+\tau g_{v})^{2}+4\varepsilon \delta q^{2}]^{1/2}}\\
                       -\frac{3c_{0}^{2}(1+\tau g_{v})^{2}\big(\delta-d(1+\tau g_{v}))}{\sqrt{2} q^{2} [c_{0}^{2}(1+\tau g_{v})^{2}+4\varepsilon \delta q^{2}]^{3/2}}.
\end{equation}
It shows that the reduced parameter $d$ is related with the control parameters $\delta$, $\varepsilon$, $a_{1}$, and the feedback parameters in the model. If $d$ is negative, the front becomes unstable upon suffering transverse perturbations resulting in the labyrinthine pattern as shown in Fig. 5 (c). If $d$ is positive, the front keeps stable to transverse perturbations leading to the spiral wave as shown in Fig. 5 (b).

For the Ising front $c_{0}$$=$$0$, so, we have:

\begin{equation}\label{}
   1-d(1+\tau g_{u}) = \frac{3\big(\delta-d(1+\tau g_{v}))}{2\sqrt{2\varepsilon\delta} q^{3}}.
\end{equation}

\indent At the critical point $d$$=$$0$, the Ising front will undergo transverse instability. So we obtain the critical line for the Ising front:

\begin{equation}\label{}
   \delta_{I}=\frac{8\varepsilon q^{6}}{9}.
\end{equation}

\indent It can be seen that the transverse instability boundary for the Ising front is unaffected by the time delay as shown by the solid line $\delta_{I}$ in Fig. 8. For the Bloch front, $c_{0}$$\neq$$0$. At the critical point to transverse perturbation ($d$$=$$0$), we obtain an implicit expression for the Bloch front:

\begin{equation}\label{}
   \frac{8}{9}q^{6} \varepsilon \delta^{2}_{BD} \frac{1+\tau g_{u}}{1+\tau g_{v}} + \delta_{BD} \left(\frac{1+\tau g_{u}}{1+\tau g_{v}}-\frac{1+\tau g_{v}}{1+\tau g_{u}}\right)=1.
\end{equation}

\indent The positive solution of $\delta_{BD}$ defines a boundary of the transverse instability of Bloch front as shown by the thin dash dot lines in Fig. 8. From Eq. (16) it is found that the competition between $g_{u}$ and $g_{v}$ alters the boundary. If $g_{u}$$>$$g_{v}$ ($g_{u}$$<$$g_{v}$), the boundary moves down (up) as shown in Fig. 8 (a) [Fig. 8 (b)]. When $g_{u}$$=$$g_{v}$ the boundary stays constant as the case without the time delay, which means that the time delay do not affect the critical stability of Bloch front to transverse perturbations if the feedback intensity $g_{u}$ equals to $g_{v}$.

\section{CONCLUSION AND REMARKS}

\indent In this work, we have studied the effects of the time-delayed feedback on the NIB bifurcation in a bistable medium. The results have shown that the time-delayed feedback applied to the activator and/or the inhibitor changes the critical point of NIB bifurcation. The time delay alters the temporal scales of the reactions, therefore the velocity of Bloch front. Large delay with weak feedback intensity is equivalent to small delay with strong feedback intensity. The effect of time-delayed feedback on the activator opposes that on the inhibitor. So there exists competition between the two feedbacks on controlling the NIB bifurcation. Upon increasing the ratio $g_{u}/g_{v}$, the critical point of NIB bifurcation shifts left which could result in a transition from Bloch front to Ising front, and vice versa. When time-delayed feedback is applied individually to one of the species, positive and negative feedback make the bifurcation point shift to different directions. In the two-dimensional case, the time delay can change the stability of front to transverse perturbations. If $g_{u}$$<$$g_{v}$, it could stabilize the front upon suffering transverse perturbation, and vice versa. In some sense applying the time-delayed feedback to species is equivalent to changing their diffusion coefficients. Thus, the wavelength of patterns can be controlled by properly using feedback parameters.

\indent Although this FitzHugh-Nagumo model is a generic model, it has described successfully the dynamics of pattern formation in bistable Ferrocyanide-Iodate-Sulfite reactions, such as the bistable spirals, oscillating spots, and labyrinthine patterns \cite{LiGe, Szalai, Lee1, Lee2}. These phenomena have been attributed to the NIB front bifurcation. In this paper, we focus on the generalized controlling scheme to the NIB bifurcation by applying time-delayed feedback to one or two of the variables. The results have shown the flexibility of this strategy on controlling the NIB bifurcation, therefore the transformation of patterns.

\indent Many real chemical experiments, such as the ferroin-, Ru(bpy)$_{3}$-, and cerium-catalyzed Belousov-Zhabotinsky systems, are sensitive to visible and/or ultraviolet light. \cite{Lee1,Lee2,Kheowan,Guo,Toth,Vanag1,Vanag2,Kheowan2,Zykov3,Hildebrand} People have realized controlling of pattern formation by the time-delayed feedback in light-sensitive chemical reactions. For example, by projecting the delayed image uniformly from the feedback loop to the gel in the Petri dish, Kheowan and Zykov realized the controlling of spiral waves in a thin layer of the light-sensitive Belousov-Zhabotinsky reaction. \cite{Kheowan2,Zykov3} The radius of the attractor for meandering spiral waves can be effectively manipulated by varying the delayed time in the feedback loop. Karl Vanag $et$ $al$ observed oscillatory cluster patterns in a light-sensitive Ru(bpy)$_{3}$-catalyzed Belousov-Zhabotinsky reaction. \cite{Vanag2} The catalyst Ru(bpy)$_{3}$ is light-sensitive. Thus, a proper illumination of the active chemical substrate can be used for spatial control of the inhibiting process (Br$^{-}$). Our results have also confirmed that applying the time-delayed feedback only to the inhibitor is enough to control the pattern formation. In a light-sensitive ferrocyanide-iodate-sulphite reaction, Lee $et$ $al$ observed the pattern transformation via NIB bifurcation by changing the flow rate or the input ferrocyanide concentration. \cite{Lee1,Lee2} Our results have shown that applying the time-delayed feedback for controlling the NIB bifurcation, from the experimental viewpoint, is equivalent to changing the residence time or the input ferrocyanide concentration. We hope that our results can be verified in one of the light-sensitive reactions with patterned (not uniform) illumination after feedback loop. The feedback loop should mainly include: 1) CCD camera, 2) video recorder, 3) computer which implements the algorithm of eqs. (3) and (4) and outputs the results (patterned images with appropriate intensity) to a projector, 4) projector which projects the patterned images inputted from the computer to the chemical substrate.

\begin{acknowledgments}
\indent This work is supported in part by Hong Kong Baptist University and the Hong Kong Research Grants Council. Y. F. He also acknowledges the National Natural Science Foundation of China with Grant No. 10975043, 10947166, 10775037, and the Research Foundation of Education Bureau of Hebei Province, China (Grant No. 2009108).
\end{acknowledgments}

\bibliography{aipsamp}

\end{document}